# What Will Remain Human in Software Architecture? A Focus Group Report

U. van Heesch, TH Köln, Germany
O. Zimmermann, University of St.Gallen, Switzerland
C. Kohls, TH Köln, Germany

**Abstract.** AI development agents are increasingly used to support and partially automate software architecture tasks. To explore how practitioners perceive this shift, specifically what changes, what remains, and what new responsibilities emerge, we conducted a focus group at the 31st European Conference on Pattern Languages of Programs, People, and Practices (EuroPLoP 2026). Twenty-two participants from industry and academia discussed current practices, trust and validation strategies, the boundaries of AI autonomy, governance challenges, and implications for education. Among others, we found broad consensus that architectural decision-making, accountability, and the authoring of architectural guardrails remain fundamentally human tasks. A central emergent concept was harness engineering: the discipline of building the system that governs AI-assisted system creation, comprising validation mechanisms, knowledge layers, and company-specific standards. The participants agreed that criticality, understood as the combination of uncertainty and cost of change, serves as the universal criterion for calibrating human oversight. A further concern was cognitive debt: the progressive erosion of human understanding of the system when AI-assisted decisions are accepted without full intellectual engagement. In this report, we present the findings of the focus group and describe directions for future work.[1]

**Keywords:** Software Architecture, AI Agents, Harness Engineering, Cognitive Debt, Focus Group, Architectural Decision-Making, Human-AI Collaboration

[1] **Preprint notice.** This is the authors' accepted manuscript of a paper accepted, after peer review, at the 31st European Conference on Pattern Languages of Programs, People, and Practices (EuroPLoP 2026). It is not the version of record. The version of record will be published in the EuroPLoP 2026 proceedings in Springer Lecture Notes in Computer Science (LNCS) and, once available, will be accessible via its DOI. Please cite the version of record.

# 1 Introduction

In the past two years, a new class of AI tools has emerged that goes beyond code completion: autonomous development agents capable of proposing, implementing, and modifying software architecture decisions. Tools such as Claude Code[2], GitHub Copilot[3], Cursor[4], and Kiro[5] operate with increasingly broad context windows and can reason about system structure, not merely individual functions. This raises fundamental questions about the future role of human software architects.

This transformation has been described from a practitioner's perspective by van Heesch [1]: the core of software engineering is shifting from implementation to intent, that is, from writing code to defining constraints, encoding quality, and validating outcomes. This shift has progressed from code completion tools through autonomous agents to what participants in our focus group described as a subcontractor relationship. Effective delegation to AI agents requires that tacit knowledge be made explicit in configuration artifacts, validation mechanisms, and architectural guardrails.

To explore how a broader group of practitioners perceives this shift, we conducted a focus group at EuroPLoP 2026. The focus group discussion guide was prepared with three exploratory interviews with experienced practitioners that were conducted prior to the focus group as part of a larger research effort. These interviews are not reported in this paper. The focus group served to explore how practitioners and researchers interpret the themes that emerged from the interviews, and to identify additional perspectives, explanations, and tensions not visible from individual experience. We chose EuroPLoP as the venue because this conference series has captured design and architecture knowledge for more than 30 years, attracting experienced practitioners and academics from multiple domains, including software architecture.

Among the central findings are the emergence of harness engineering, the discipline of building the system that governs AI-assisted system creation, as a new field of practice, and cognitive debt as a systemic risk when practitioners delegate without full intellectual engagement. Validation of AI-generated artifacts emerges as the key challenge across all levels of architectural work.

This report describes the setup, participants, and procedure of the focus group, presents and discusses the findings, and concludes with directions for future work. It reports focus group experience rather than results of a formal empirical study. The focus group data will also feed into a larger study that combines the exploratory interviews with the focus group findings.

The rest of this paper is organized as follows: Section 2 provides background and positions our work relative to existing literature. Section 3 explains the setup of the focus group and the characteristics of the participants. Section 4 presents the results of the discussion, structured by the discussion topics. Section 5 discusses cross-cutting themes and tensions. Section 6 describes limitations and threats to validity, and Section 7 summarizes and presents directions for future work.

[2] https://docs.anthropic.com/en/docs/claude-code
[3] https://github.com/features/copilot
[4] https://www.cursor.com
[5] https://kiro.dev

# 2 Background and Related Work

Software architecture is concerned with the fundamental structural decisions of a system, including component decomposition, quality attribute trade-offs, and long-term evolution [2]. As AI agents increasingly participate in these activities, several bodies of related work become relevant.

## 2.1 Architecture Decisions

Software architecture decisions have been studied extensively. Tyree and Akerman [3] proposed structured decision templates with explicit rationale, assumptions, and consequences. Jansen and Bosch [4] presented a model for architecture decisions as first-class entities with dependencies between them. Van Heesch et al. [5] proposed a documentation framework that integrates decisions across multiple architectural views and supports reasoning about decision dependencies and evolution. Zdun et al. [6] investigated sustainable architectural design decisions and their long-term viability. In parallel, Nygard [7] introduced lightweight Architecture Decision Records (ADRs) as a practical means of capturing individual decisions in short, version-controlled text files. Keeling [8] further popularized ADRs as part of a broader design toolkit for practicing architects. While the richer frameworks discussed above [5], [6] provide deeper support for reasoning about trade-offs, alternatives, and long-term consequences, ADRs in the style of Nygard have gained popularity in agile teams due to their low adoption barrier, though at the cost of reduced expressiveness for decision scenarios with medium or high complexity.

More recently, Zimmermann and Stocker [9] consolidated selected architecture decision practices within their Design Practice Reference, including a definition of done. The Markdown Architectural Decision Records (MADR) project [10] builds upon earlier work on templates and suggests both a comprehensive and a minimal template for decision capturing.

Both lightweight and comprehensive approaches are relevant to our findings, as participants reported using AI to generate ADRs while insisting on human judgment for the underlying decision rationale.

## 2.2 Human-AI Collaboration in Software Engineering

Recent empirical work has begun to investigate how developers interact with AI coding assistants. Barke et al. [11] studied GitHub Copilot users and identified two interaction modes: acceleration (using AI to speed up known tasks) and exploration (using AI to investigate unfamiliar domains). Liang et al. [12] surveyed professional developers and found that trust in AI code suggestions increases with experience but remains calibrated by task complexity.

Hassan et al. [13] proposed foundational pillars for agentic software engineering, framing the field as a research discipline in its own right. He et al. [14] found that Cursor AI increases short-term development velocity but also long-term code complexity, suggesting that AI-assisted productivity gains may come at the cost of maintainability. These findings provide context for understanding how practitioners calibrate trust in AI architecture proposals specifically.

## 2.3 AI for Software Architecture

Bucaioni et al. [15] conducted a peer-reviewed systematic literature review on AI support for software architecture practice, synthesizing 51 primary studies and mapping them against 17 practitioner-reported challenges. They identify six AI-specific challenges, including the need for context-aware reasoning, incorporation of domain-specific expertise into AI-supported reviews, and integration of long-term evolution and debt management into AI recommendations (which they label AICH6). Their proposed roadmap envisions human architects as "chief strategists" and AI as "chief analyst," but acknowledges that current approaches offer little guidance on how this collaboration should work in practice.

Our focus group complements the literature-based analysis with direct practitioner evidence from 22 participants actively using AI agents in architecture work. Where Bucaioni et al. identify gaps in the literature, our participants describe emergent practices that address these gaps: harness engineering as a concrete governance model, cognitive debt as the human dimension of their AICH6, and detailed accounts of how trust calibration works in daily practice.

## 2.4 AI Configuration and Context Engineering

A growing body of work investigates how developers configure and steer AI agents. Galster et al. [16] identified a taxonomy of eight configuration mechanisms (context files, skills, rules, hooks, subagents, commands, settings, and MCP configurations) through which developers tailor agentic tool behavior to their projects. Mohsenimofidi et al. [17] studied context engineering practices across open-source repositories, analyzing how developers encode project constraints and architectural decisions in configuration artifacts.

Van Heesch [1] described from a practitioner's perspective how tacit engineering knowledge must be translated into agent configuration, reusable skills, and automated validation mechanisms when agents become operational actors in development teams. These configuration artifacts correspond closely to what our focus group participants described as the "harness": version-controlled, human-authored specifications that govern AI agent behavior within a project.

### 2.5 Automated Architecture Governance

The idea of encoding architectural constraints in executable form has a growing tradition within the architecture community. Ford et al. [18] introduced architectural fitness functions as automated checks that protect architectural characteristics as systems evolve, analogous to unit tests for code. Tools such as ArchUnit[6] and jQAssistant[7] enable architecture conformance checking by expressing structural rules as code that runs in CI/CD pipelines. These approaches share a common principle: making architectural intent explicit, machine-readable, and automatically enforceable.

Our focus group revealed an emerging discipline, harness engineering, that extends this principle to a new domain: governing not the system itself but the AI-assisted process by which the system is created.

### 2.6 Community Discussions on the Future of Software Architecture

Taibi et al. [19] organized a Birds-of-a-Feather session at ICSE 2026 to identify "big questions" in software architecture research. Their findings overlap substantially with our focus group results: participants discussed the evolving role of architects in AI-driven systems, cognitive debt in AI-assisted development, the need for human validation of AI-generated decisions, and the question of when architectural intervention is required.

Their open research problems, including loss of human control, architecture knowledge loss, and certification of AI-assisted decisions, confirm several concerns raised independently in our focus group. Our work complements theirs by providing richer practitioner narratives and by identifying harness engineering as a concrete emergent discipline that addresses several of these challenges.

## 3 Focus Group Setup and Participants

This section specifies the organizational preparation and setup of the focus group, as well as the participant information (demographics, experience with AI).

### 3.1 Setup

The focus group took place at EuroPLoP 2026 in Irsee, Germany, in July 2026. We announced the session prior to and during the conference. Participation was voluntary and advance registration not required. The session lasted 90 minutes and was held in a co-located, circular seating arrangement. The research team consisted of a primary moderator, a secondary moderator who monitored question guide coverage and balanced participation, and a third researcher.

[6] https://www.archunit.org
[7] https://jqassistant.org

All three researchers took notes during the session and complemented them immediately after. The discussion was audio-recorded with the explicit informed consent of all participants.

With 22 participants, the session exceeds the typical focus group size of 6–12 participants recommended by [20]. In practice, the session functioned as a moderated plenary discussion rather than a classical focus group with breakout sessions. The moderator actively invited contributions from quieter participants and ensured that dominant voices did not monopolize the discussion. Based on the researchers' notes, 19 of the 22 participants contributed actively, though at varying levels of intensity.

We structured the discussion around six thematic parts derived from the preceding exploratory interviews:

- Part A: Current use of AI tools for architecture tasks
- Part B: Trust and validation of AI proposals
- Part C: Tasks that remain fundamentally human
- Part D: New responsibilities and governance
- Part E: Autonomous decisions and accountability
- Part F: Implications for education and skill development

For each part, we prepared lead questions, follow-up questions, probes, and stimulus scenarios drawn from the interview findings. In practice, the discussion flowed naturally between topics, and not all prepared stimuli were needed. Parts C and D generated the most extensive discussion and partially absorbed topics planned for Part E.

### 3.2 Participants

In total, 22 participants took part in the session, in addition to the three members of the research team. Prior to the session, participants completed a short anonymous questionnaire (collected separately from the signed informed consent form). Eighteen participants returned the questionnaire (response rate: 82%). The participants came from both industry and academia: twelve held academic positions (professors, lecturers, researchers, PhD candidates) and six held industry positions (CTOs, developers, team leads, project managers).

Professional experience in software development ranged from 3 to 30 years (median: 14, mean: 13.2). Architecture experience ranged from 0 to 23 years (median: 5, mean: 6.7). Self-rated architecture expertise was distributed as follows: Expert (1), Proficient (4), Competent (7), Advanced Beginner (5), Novice (1).

Regarding AI tool usage, ChatGPT (14) and Claude (13) were the most widely used tools, followed by GitHub Copilot (8) and Cursor (3). Seven participants had been using AI tools for 6–12 months, five for 1–2 years, and two for more than 2 years. Usage frequency varied: three participants reported continuous use for nearly all professional tasks, three used AI daily, three several times per week, four weekly, and five occasionally. Self-rated AI tool expertise was predominantly intermediate (8) or novice (6), with two experts and one advanced user; one participant did not answer this question.

Participants worked primarily on Web/Cloud applications (13), distributed systems (7), data-intensive systems (7), enterprise systems (7), and embedded systems (6). Team sizes were predominantly small: fewer than 5 (9) or 5–15 (7).

### 3.3 Data Collection and Analysis

The discussion was audio-recorded and transcribed locally using faster-whisper (v1.2.1, based on OpenAI Whisper Medium, 769M parameters) with Silero VAD filtering. The language was set to English manually. The resulting transcript was manually corrected for technical terms and proper nouns by the first author. Individual speaker attribution was not performed; only moderator contributions were marked.

The analysis drew on three sources: the corrected transcript, notes taken by the researchers during the session, and notes complemented immediately afterwards. The first author structured the findings from the transcript, organized by the six thematic parts. All three researchers validated the assignment of participant statements to themes and checked the reported results against their own session notes. Disagreements were resolved through discussion. The analysis is a structured synthesis of the discussion rather than a formal qualitative coding procedure. To validate the findings, we conducted member-checking with five participants who had indicated during the focus group that they wished to be informed about the results: they reviewed the findings as reported in this paper and confirmed them. Participant quotes have been lightly edited for readability (removal of filler words and false starts) without altering their meaning. The informal, conversational tone of the discussion has been preserved to maintain authenticity.

## 4 Results

This section is structured according to the six discussion parts introduced in the specification of the setup in Section 3.

### 4.1 Current Use of AI Tools for Architecture Tasks (Part A)

The opening discussion revealed that AI tools are used broadly across architecture-related tasks. Documenting architecture decisions, particularly ADRs, was mentioned by multiple participants as one of the most common and uncontroversial uses. A closely related effort is gathering and preparing context for the coding agents and the harness; one participant summarized both as "the tedious job of documenting and gathering the context for the coding agents and harness."

Others confirmed using AI for C4 architecture documentation [21] and for maintaining decision records as the codebase evolves. These findings match the AI usage in practice reported in [22].

Beyond documentation, participants reported using AI tools for brainstorming and exploring architectural alternatives. One practitioner described with enthusiasm how Claude not only confirmed that a component should be extracted into its own service, but immediately implemented the change: "The damn thing went and split the whole thing into a separate service after a few back and forth." This illustrates how AI has moved beyond advisory roles into active implementation of architectural changes.

A recurring theme was the importance of detailed upfront specification before engaging AI agents for implementation. One participant explained: "whenever I just give it a one-line prompt, it usually starts doing weird stuff, because it assumes things that I thought a different way. So, I usually start with a very detailed implementation plan." This *specification first* approach was echoed by others who found that the quality of AI output depends directly on the precision of the input specification.

Participants also observed that AI agents struggle when starting from an empty repository. One practitioner explained that "I first have to implement one or two feature slices by myself, because otherwise it takes me too much time to learn the agent how that code should look like." AI works more efficiently with existing code as a reference to establish a desired architectural style.

Additional use cases included exploring unfamiliar technology stacks, performing security threat analysis with frameworks such as STRIDE [23], analyzing legacy systems to reverse-engineer architecture documentation, and generating architectural prototypes. One participant noted that structured analysis tasks are particularly well-suited to AI because of their systematic nature.

An interesting finding concerned the dual role of ADRs in AI-assisted development. Beyond their traditional function as human documentation, ADRs now serve as persistent memory for AI agents across sessions. One participant observed: "They have a memory of a one-day fly or they lack that memory, so these design decisions also help solidify the architecture that you're in."

When asked about deliberate non-use of AI, the group revealed a clear divergence. Some participants deliberately avoid AI for certain tasks to preserve their own skills: "There's lots of things that I don't let the AI do, because I fear that I will lose the skills myself, and of course I enjoy doing it myself." Others took the opposite stance: "Why would I do something that the thing can do itself?" This tension between skill preservation and efficiency remained unresolved throughout the session.

## 4.2 Trust and Validation of AI Proposals (Part B)

The dominant mental model for AI trust was that of a junior colleague or subcontractor. One participant stated: "I do a technical analysis as if it's going to be a colleague suggesting an architecture." Another agreed explicitly with an outsourcing analogy: "It's the same as if you have outsourced software [...]. I very much agree with the outsourcing thing. I have the same feeling about AI."

The group reached strong consensus that trust in AI proposals is calibrated by the criticality of the decision or task at hand. Low-consequence decisions can be accepted without full understanding; high-consequence decisions require thorough verification. One participant formulated this principle concisely: "I can accept things that I don't fully understand if I can estimate that the blast radius is more or less [bounded]." In other words, plausibility suffices when the stakes are low. The proximity to production emerged as a key factor: "The closer I go to production, the closer I have 10 million users in a database, the more scared I am." When the decision is consequential, practitioners experiment to validate.

A notable finding was the consensus that existing software engineering validation mechanisms remain the appropriate quality backbone. One participant stated: "The SonarQube[8] harness – nothing changed between when humans coded and made mistakes and now when an AI does. It's the same level of distrust." CI/CD pipelines, automated tests (from unit tests to end-to-end tests), tool-supported code quality analysis and similar tools apply equally to AI-generated code and to human-written code.

One participant introduced the practice of using multiple models for cross-validation, a technique related to the LLM-as-a-Judge paradigm [24]: "Sometimes I throw the requirements between models around and watch them insult each other." Using one model to review another's output was presented as a lightweight quality assurance technique.

## 4.3 Tasks that Remain Fundamentally Human (Part C)

The discussion about fundamentally human tasks generated the broadest and deepest engagement. The group discussion reached strong consensus on three areas: architectural decision-making, accountability, and the authoring of guardrails.

Regarding decision ownership, one participant stated: "I think what should remain human is the overall architecture, the division of responsibilities, components, and how you organize, because it can be AI-aided, but the human needs to make the decisions, because this is highly strategic on how you evolve the architecture." AI assists with preparation (generating options, evaluating trade-offs, exploring alternatives), but the final decision remains with the human architect.

[8] https://www.sonarsource.com/products/sonarqube/

The authoring of architectural rules was identified as an exclusively human task. One participant reported: “This is one document we currently say only humans can touch. It’s given the AI as a background to develop things, but it’s a document… No change through AI, it needs to be written by human because it needs to be precise, short, and that’s ruling everyone, the human and the AI works with.” Examples included technology stack constraints such as mandating Kubernetes, or development culture norms, all of which are strategic decisions whose rationale AI cannot assess. The rules document governs both human developers and AI agents, and its precision cannot be delegated. This extends to organizational context more broadly: AI does not know team skills, strategic goals, or company constraints, and initial system decisions require human judgment to set these baselines.

Accountability emerged as a non-negotiable human responsibility. One participant framed this definitively: “It should never be ‘the AI did it’. It should always be, I did it, and I let the AI do the practical work, and I’m responsible for what the AI did, because it was my task to check it, validate it, and I’m standing here before the body and explain that I decided that this got to production.” Another participant, approaching the question from a legal perspective, added: “As a lawyer, what still remains human? I need one human who is accountable and who I can punish.” However, as discussed in Part E, participants also questioned whether individual accountability scales to large AI-managed systems.

A nuanced contribution concerned the risk of solution space blindness. One participant argued that AI provides a working solution but may prevent architects from exploring the full solution space: “The solution space is actually much bigger, and I’m blind to it. If I didn’t do the digging into the topic, I’m blind to X percent of the solution space. That doesn’t make me a bad architect in terms of accountability – I’ve submitted a proper solution, I’m just not brilliant anymore.”

The skill degradation discussion produced a compelling analogy from aviation: “Aviation authorities demand that pilots fly a certain amount of time without autopilot but themselves. It goes back to the incident that we had in the middle of the Atlantic because the young pilots would not fly enough anymore.” The group partially agreed that programming skills are more at risk of degradation than architectural thinking skills, since architecture has always been about decisions and quality trade-offs rather than the mechanical act of translating specifications into code.

Related to skill degradation, the concept of cognitive debt was introduced as a companion to technical debt [25]: “There used to be technical debt that most architects have heard. Now it’s cognitive debt. When you interact with an AI and you do some stuff and you lose the knowledge of what happens there. You incur no longer the technical debt but the cognitive debt.” Where skill degradation concerns the loss of a previously held ability, cognitive debt describes the conscious or gradual acceptance of not fully understanding parts of the system one is responsible for. Over time, this erodes an organization’s ability to solve future problems.

An observation with structural implications was that good modular architecture enables selective human oversight. If a component is well-contained and its properties are observable and measurable from the outside, an architect does not need to understand its internals: "I don't see as a loss of responsibility the fact that I don't know inside anymore, if it's contained, and I can measure and see what are the properties I have." This means the architect's role shifts towards deciding where to invest attention, based on the criticality of individual components as well as its interface contract/specification.

The group also noted the loss of traditional mentoring patterns: "And now what happens is that instead of going to talk to the other person, you ask the AI. So this link, it was important to pass this experience. It is lost." The knowledge transfer chain between senior and junior practitioners is being disrupted when juniors turn to AI rather than to experienced colleagues.

## 4.4 New Responsibilities and Governance (Part D)

A central emergent concept of the session was harness engineering. One participant introduced the term and the group built upon it extensively. The harness was defined as "everything that's not the LLM and not the chatbot. Everything else is the harness." More precisely: "Everything you put in place to validate, verify the output of the LLM, all the instructions about context and what has been decided in this company, goes in there, goes both specification and validation."

The framing of harness engineering as "the system that builds the system" captured the group's imagination: "We are building a totally new layer. We are building the factories, how companies build their systems. The system that builds the system – we are building that automation layer. It's totally novel." This represents a meta-architectural discipline: designing the constraints, validation mechanisms, and knowledge layers that govern AI-assisted system creation.

The group identified key components of a harness: LLM-agnosticism as an architectural pattern, a knowledge layer encoding company standards, an observability layer, and hybrid validation mechanisms mixing LLM-based checks with deterministic ones. Companies that reported positive results were those that trained their harnesses on their own definition of good software development. This observation reinforces the group's earlier message that existing software engineering validation mechanisms remain the appropriate quality backbone.

A critical gap was identified: quality assurance of the harness itself. One participant stated bluntly: "Nobody even knows how to do quality assurance of that asset. How do you have quality assurance practices on that level, for example? We only are able to see from the output the artifacts that is this good or bad, but not on that level." The discipline lacks established practices for evaluating whether a harness is well-designed. In current practice, harness quality can only be assessed indirectly, by judging the generated output. This is an unreliable proxy: good output does not confirm a sound harness, since it may result from a simple task or a capable model rather than the harness itself. A direct, output-independent evaluation of harness quality remains an open research problem.

A related concern was harness evolution. One participant asked: "How do you evolve the harness over time, over changing the environment?" The group acknowledged that a harness is not a one-off artifact but requires continuous maintenance as both AI capabilities and organizational standards evolve.

One participant mentioned a story about a concrete case that illustrates the potential of the harness approach: a logistics company with a multi-million-line legacy backend that was considered impossible to modernize was reportedly reduced to a functionally equivalent system of significantly smaller size by a small team that spent several weeks building a harness to understand the system's invariants rather than its code. This story is unverified, but it illustrates the kind of potential that participants attribute to the harness approach.

The discussion revealed a vision of a two-type software world emerging from AI-assisted development. One participant distinguished between "demo software and quality software. Demo software will be everywhere." Non-engineers using tools such as Lovable[9] create applications directly, without professional involvement. For simple applications, "the users will do it themselves." This distinction resonates with Meyer's ABC classification of software criticality [26], [27]: the participants' "demo software" corresponds to casual (C) applications, while "quality software" maps to business-critical (B) and acute (A) systems. However, the group expressed strong concern about this boundary blurring: "anything that I deem where trust in the application is necessary, I am very scared of that future. That my banking app will be vibe-coded. That my medical app will be vibe-coded." The engineering effort, several participants noted, is shifting from writing application code to building and maintaining the harness: "Maybe in this case the whole engineering effort and actual work will go into the harness."

## 4.5 Autonomous Decisions and Accountability (Part E)

The discussion about autonomy overlapped substantially with Parts C and D. The group's position on accountability was clear: a human must always be accountable. One participant drew on autonomous driving analogies, noting that legal frameworks require an identifiable person who can be held responsible.

However, making a single person accountable for a large AI-managed system was questioned: "If it really comes out, a single person on a huge system will not be accepted. Either it's accepted that AI can take over and that the company that delivers AI has to compensate for the damage, or you have to have an appropriate amount of staff that can really oversee your system." Two alternative models emerged: vendor liability (analogous to autonomous vehicles) or maintaining sufficient oversight staff.

[9] https://lovable.dev

The expanding span of architect responsibility was discussed as a natural consequence of AI tools. One participant framed this as the latest step in an ongoing abstraction: "You will always go to a level, the next level of abstraction, and then you are there responsible." A single architect can now manage larger system portions in terms of design, development, and evaluation speed.

An important structural insight emerged regarding vibe coders who bypass explicit architecture: "the vibe coders, they skip architecture. They go directly from roughly specified requirements. But the harness does it. So whoever did the harness, did the architecture." Architecture does not disappear when end users or non-engineers create applications; it shifts to the platform and harness level.

## 4.6 Implications for Education and Skill Development (Part F)

The education discussion produced consensus that computer science fundamentals remain essential: "They still have to learn how the software works. As long as it's there. And even if it's on the line and hidden behind layers of AI and generated stuff. They still need to understand." Computer science and software engineering fundamentals such as programming, operating systems, and networking must continue to be taught. The curriculum expands rather than contracts.

Core architectural competencies and principles such as modularity, separation of concerns, quality attributes, and trade-off analysis remain essential curriculum content. AI tool usage is additive, not a replacement. One participant emphasized that students "need to be able to confidently make decisions. And they need the knowledge to make those decisions confidently."

One area where curriculum may be reduced is the breadth of programming language exposure: "The variety of programming languages is not that necessary anymore. And what becomes necessary is many things that we talked about today. How to build up these harnesses. How to make quality assurance better. How to learn to exactly specify things." Harness engineering, specification precision, and quality assurance (validation) replace multi-language breadth.

Participants also noted that the new discipline provides its own curriculum filter. One participant argued that the requirements of harness engineering can serve as a guide for what to prioritize: "What is important in harness engineering is going to be a filter towards the skills that you don't have to learn anymore." The core new competencies that emerged were defining precise inputs for AI agents and understanding their outputs, as well as the recognition that maintenance of both systems and harnesses will continue to matter.

A mathematics analogy resonated with the group: “I think we should systematically look at the transformation of math. Because that’s from my perspective the last field that had such a fundamental transformation. And there are still mathematicians out there.” The mechanical computation role (human calculators) was eliminated by computers, but the conceptual and creative role persisted. Similarly, routine coding may be automated while architectural thinking remains human.

The discussion concluded with the observation that resilience may be the most important meta-skill to cultivate in students: “We don’t know what’s going to be five, ten years from now. But they have to be ready for anything. We don’t want them to feel like they should give up. Resilience is the biggest thing.” This implies that beyond technical skills, curricula must foster the confidence to navigate uncertainty. Students need the ability to remain effective when established practices are disrupted, to critically evaluate new ways of working rather than passively adopting them, and to return to first principles when automation produces results they cannot explain.

# 5 Discussion and Reflection

This section interprets the findings from the previous section. We organize the discussion around cross-cutting themes that emerged across multiple parts of the focus group, position harness engineering as an emergent discipline, and examine tensions that remained unresolved.

## 5.1 Cross-Cutting Themes

Several themes emerged across multiple parts of the discussion.

**Criticality as the universal decision criterion.** Throughout the session, criticality served as the axis along which all decisions about AI autonomy were calibrated. Low criticality permits more AI autonomy and less human verification; high criticality demands human control and full understanding. This applies at multiple levels: individual decisions, entire systems, and even the choice of the development approach. One participant articulated criticality as the combination of uncertainty and cost of change, a framing that was accepted by the group without dissent and that resonates with risk-driven architecture principles [5], [28], [29].

**The outsourcing metaphor.** The dominant mental model frames AI as a subcontractor or outsourced team member. Trust, verification, and management patterns transfer directly from existing professional relationships. This framing normalizes AI interaction and implies that fundamentally new management approaches may not be necessary; rather, existing ones should be applied with appropriate rigor.

**Specification as the primary human artifact.** As code generation becomes automated, the specification becomes the central artifact through which humans contribute. Precise specification is both the input that controls AI quality and the primary skill architects develop. Multiple participants reported that working with AI has improved their specification skills, because imprecision leads directly to wrong outputs.

**Historical continuity and genuine novelty.** The group oscillated between two narratives: "nothing fundamentally changed, we apply the same principles at a higher level" and "this is totally novel, a new layer nobody knows how to quality-assure." Both appear to be simultaneously true: the principles of sound engineering persist, but the system structure and where architectural decisions are made are genuinely new.

**The Mythical Man-Month revisited.** Brooks [30] observed that adding people to a late software project makes it later, because communication overhead grows quadratically with team size. One participant enthusiastically claimed that AI invalidates Brooks's Law by reducing the need for large teams and inter-team communication. A single architect with AI support can achieve what previously required a team. However, the immediate response ("tell that to the enterprise") captured widespread skepticism about whether large organizations will adopt this model.

**Cognitive debt as a systemic risk.** A participant introduced the term cognitive debt as a counterpart to technical debt: the organizational knowledge loss incurred when AI handles tasks that humans previously performed with full cognitive engagement. Our interpretation extends this observation. Cognitive debt accumulates through three mechanisms. First, practitioners accept AI-proposed decisions without fully understanding their rationale, particularly when the decision appears plausible and the blast radius seems bounded. Second, even when practitioners briefly engage with a decision, the reduced duration and depth of that engagement means the knowledge is not sufficiently anchored in memory. Third, what we call fly-by approval: when AI agents request confirmation frequently, practitioners develop a pattern of routinized consent without conscious evaluation, analogous to dismissing cookie consent banners. The approval is given, but no cognitive processing takes place. Over time, these mechanisms erode the architect's intellectual model of the system as a whole. Where previously an architect held a coherent mental representation of the system's structure, constraints, and trade-offs, AI-assisted workflows fragment this understanding into shallow, episodic interactions. The concept connects directly to the plausibility-is-enough stance discussed in Part B, and to the skill degradation and solution space blindness concerns in Part C. Cognitive debt can be seen as a counterpart to the well-known phenomenon of architectural knowledge vaporization [4], but driven by insufficient cognitive engagement rather than insufficient documentation.

## 5.2 Harness Engineering as Emergent Discipline

As previously reported, a central emergent concept of the focus group was harness engineering as a new architectural discipline. The term has been independently described by Böckeler [31] as the practice of building guides and sensors that increase the probability of good agent outputs. We define harness engineering as this:

> Harness engineering is the design, implementation, and maintenance of the socio-technical system that governs AI-assisted software creation.

A harness encompasses all components between the human and the LLM that control context, enforce constraints, validate output, and encode organizational standards. It includes knowledge layers, observability mechanisms, deterministic validation rules, and company-specific quality criteria. The scope and the detailed design of harnesses are driven by the criticality and type of software built.

Harness engineering is distinct from DevOps, which focuses on the delivery pipeline for application code. It is also distinct from architecture conformance checking (e.g., fitness functions, ArchUnit), which validates properties of the built system against architectural rules. Harness engineering operates one level above: it governs how systems are built by AI agents rather than checking properties of the resulting artifacts. Architectural fitness functions validate the output; the harness governs the process that produces it. Empirical evidence for the emergence of this discipline can be found in recent mining studies: Galster et al. [16] identified over 15,000 configuration artifacts across nearly 5,000 repositories, and Mohsenimofidi et al. [17] documented how developers encode project constraints into context files, both of which represent concrete manifestations of harness engineering in practice.

Harness engineering connects several threads of the discussion:

- the shift of engineering effort from application code to the governance layer;
- the encoding of company knowledge into reusable constraints;
- the need for LLM-agnostic architectures; and
- the observation that "whoever did the harness, did the architecture."

The harness is where architectural decisions live when individual applications become disposable.

## 5.3 Unresolved Tensions

The discussion revealed several tensions for which the group did not reach resolution:

- **Efficiency versus skill preservation:** Maximizing AI usage risks skill degradation, but restricting usage sacrifices productivity.

- **Acceptance without full understanding versus insistence on deep comprehension:** Some practitioners accept AI proposals when the risk is bounded, others insist on understanding every decision fully.
- **Maintainability in an era of regeneration:** If AI can rewrite entire systems, does maintainability of code still matter? The instinctive answer is yes, but the intellectual argument requires more nuance. Maintainability may shift to the harness level.
- **Individual brilliance versus adequate solutions:** AI provides working solutions but may prevent exploration of the full solution space.

These tensions share a common structure: in each case, a short-term productivity gain stands in opposition to a long-term capability that is difficult to recover once lost. The group did not resolve these tensions. Instead, the recurring position was that criticality determines where on each spectrum a given team or project should operate – the same principle that governs trust calibration and the scope of human oversight throughout this report. What remains unclear is whether these tensions are transitional, resolving themselves as tools mature and practitioners adapt, or permanent features of AI-assisted architecture work. The focus group data does not answer this question, but the strength of opinion on both sides suggests that these are genuine dilemmas rather than problems awaiting a single correct solution.

# 6 Limitations and Threats to Validity

We discuss threats to validity along four standard dimensions.

## 6.1 External Validity

The participant group, while experienced and diverse in professional background, was recruited from a single conference community. EuroPLoP attendees tend to be architecture-aware, pattern-oriented, and reflective about their practice. This may not represent the broader population of software practitioners, particularly those less engaged with architecture as a discipline or those in domains where AI adoption is less advanced. The findings should be interpreted as perspectives from an expert community rather than as representative of the profession at large.

## 6.2 Internal Validity

The focus group format inherently reflects the influence of vocal participants. While the moderator actively invited quieter voices, some perspectives may be over-represented. The concept of harness engineering, for instance, was introduced primarily by one vocal participant, though it was broadly accepted and built upon by others. With 22 participants in a plenary format, individual contributions were necessarily brief, and some topics could not be explored in depth.

The session was limited to 90 minutes, and several topics could not be explored fully. Parts C and D received disproportionately more time than Parts E and F.

### 6.3 Construct Validity

The discussion guide was derived from three exploratory interviews. While this grounding in prior empirical data strengthens the relevance of the questions, it also means that the discussion was pre-structured around themes identified by a small number of cases. Topics not present in the interviews may have been underexplored.

### 6.4 Reliability

The analysis was conducted by three researchers based on the transcript and session notes. It represents a structured synthesis rather than formal qualitative coding. Member-checking was conducted with five participants who confirmed the reported findings, which strengthens the credibility of our interpretation. The findings are nevertheless best understood as the authors' interpretation of the discussion, supported by methodological transparency about the analysis process.

There is overlap between the research team and the facilitation team. The moderator's framing of questions and follow-ups inevitably influenced the direction of the discussion.

## 7 Concluding Summary and Future Work

The focus group identified several central findings. Architectural decision-making, accountability, and the authoring of guardrails remain fundamentally human responsibilities, even as AI agents take over increasing portions of implementation work. Harness engineering emerged as a new discipline: the design and maintenance of the socio-technical system that governs AI-assisted software creation, encompassing validation mechanisms, knowledge layers, and organizational standards. Cognitive debt was identified as a systemic risk: the progressive erosion of human understanding when AI-assisted decisions are accepted without full intellectual engagement. Across all themes, criticality in terms of uncertainty and cost of change served as the universal criterion for calibrating the degree of human oversight. A consequence of this principle is the emergence of a two-type software world: low-criticality applications that can be built with minimal human involvement, and trust-critical systems that demand rigorous professional engineering.

Several directions for future work emerge from these findings. Validation is key to success and harder to achieve for architecting tasks than for coding tasks. Open questions for harness engineering include:

- What are the components and architectural patterns of effective harnesses?

- How can harness quality be evaluated independently of the output it produces?
- How do organizations evolve their harnesses as AI capabilities change?
- What is the relationship between harness engineering and established approaches such as architectural fitness functions, conformance checking, and DevOps?

More broadly, how to evaluate, test, and evolve "the system that builds the system" is a question that current software engineering practice has not yet answered.

This report is an experience report, not a formal study. A qualitative analysis of the focus group data is planned, using structured thematic coding, researcher validation, and triangulation with the preceding exploratory interviews. The results will be published as a separate study report.

The tension between efficiency and skill preservation, and the concept of cognitive debt, suggest a need for longitudinal studies investigating how architectural competence evolves in practitioners who rely heavily on AI agents over extended periods. Further open questions raised but not resolved during the session include: how accountability should be distributed when AI-managed systems grow beyond what a single human can oversee; how trust-critical applications can be protected in a world where vibe-coded software is everywhere; how the traditional path from junior to senior architect can be maintained when mentoring is replaced by AI interaction; and how architects can ensure that AI does not limit them to a single solution when the actual solution space may be much larger.

Despite these open questions, the focus group's collective sentiment was clear. As one participant put it: software development is dead – long live harness engineering.

## Acknowledgments

We thank all participants of the focus group for their active engagement and willingness to share their experiences and perspectives. We also thank the EuroPLoP 2026 organisers for providing the venue and time slot. The work of Olaf Zimmermann has received funding from the Swiss National Science Foundation under Grant No. 10002384 (TaSSAreCt project).